\documentclass[12pt]{article}
\usepackage{epsfig}
\usepackage{pst-plot,url,epsf}
\newcommand{\beq}{\begin{equation}}
\newcommand{\eeq}{\end{equation}}
\newcommand{\bea}{\begin{eqnarray}}
\newcommand{\eea}{\end{eqnarray}}
\newcommand{\beas}{\begin{eqnarray*}}
\newcommand{\eeas}{\end{eqnarray*}}
\newcommand{\nn}{\nonumber}

\newcommand{\ra}{\rightarrow}

\newcommand{\AmS}{{\protect\the\textfont2
  A\kern-.1667em\lower.5ex\hbox{M}\kern-.125emS}}

\begin{document}
\thispagestyle{empty}
\begin{flushright}
October 2013\\
\vspace*{1.5cm}
\end{flushright}
\begin{center}
{\LARGE\bf The top--Higgs coupling at the LHC\footnote{
Presented at the XXXVII International Conference of Theoretical 
Physics, `` Matter to the Deepest'', Ustro\'n, Poland, September 1--6, 2013.}}\\
\vspace*{2cm}
K. Ko\l odziej\footnote{E-mail: karol.kolodziej@us.edu.pl}\\[1cm]
{\small\it
Institute of Physics, University of Silesia\\ 
ul. Uniwersytecka 4, PL-40007 Katowice, Poland}\\
\vspace*{4.5cm}
{\bf Abstract}\\
\end{center}
The factorization scale dependence of the anomalous top-Higgs
coupling effects in the leading order differential cross sections and
distributions of the secondary lepton 
in the process of associated production of the top quark pair and the Higgs 
boson at the LHC is discussed. 
It is also shown that the differential cross section
as a function of the rapidity of the secondary lepton 
in the process is practically not sensitive
to a sign of the anomalous pseudoscalar coupling.

\vfill

\newpage

\section{Introduction}

\begin{figure}[htb]
\centerline{
\epsfig{file=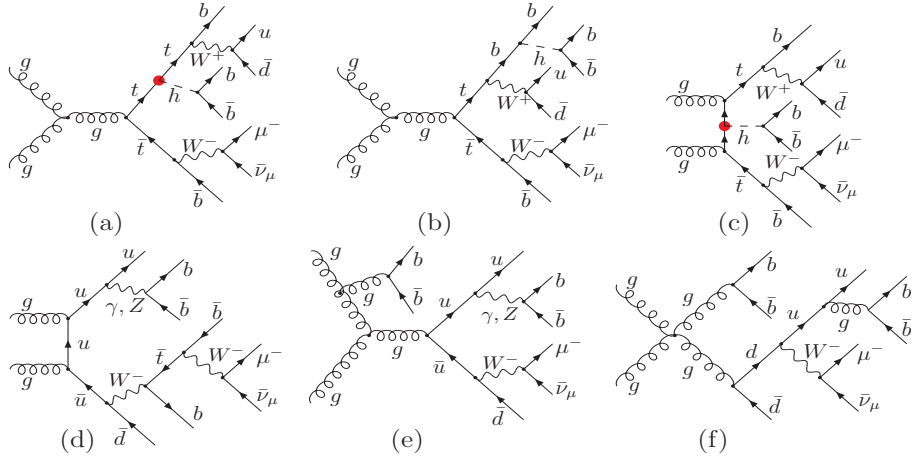,  width=120mm, height=60mm}}
\caption{Examples of the lowest order Feynman diagrams of process
(\ref{gg8f}): (a), (b) and (c) are the signal diagrams of $t\bar t h$ 
production, (d), (e) and (f) are the $t\bar t h$ background contributions. 
Blobs indicate the top--Higgs coupling.}
\label{diags}
\end{figure}

If the new particle with mass of about 125~GeV discovered at the LHC is
indeed the Higgs boson of Standard Model (SM) then practically the only model 
independent way to 
constrain its coupling to the top quark is to measure the process 
of associated production of the top quark pair and Higgs boson.
First observation of the process
\bea
\label{pptth}
pp \;\ra\; t \bar t h 
\eea
was already reported by the CMS collaboration \cite{tthCMS}.
At the LHC, process (\ref{pptth}) is dominated by the gluon fusion 
mechanism. If the dominant decay modes:
$h\to b\bar b$, $t\to bW^+$, $\bar t\to \bar bW^-$
and the subsequent decays of the $W$-bosons are taken into account
then the hard scattering partonic processes such as
\bea
\label{gg8f}
gg \;\ra\; b u \bar{d} \;\bar b \mu^- \bar \nu_{\mu} b \bar b
\eea
that corresponds to one of the $W$-bosons decaying hadronically and the other 
leptonically should be considered. Already in the leading order (LO) of 
the SM, the matrix element of process (\ref{gg8f}) in the 
unitary gauge, if calculated with the unit Cabibbo--Kobayashi--Maskawa mixing 
matrix and neglecting masses of particles lighter than 
the $b$-quark, receives contribution from $67\,300$ Feynman diagrams
some examples of which are shown in Fig.~\ref{diags}. The diagrams depicted in 
the first row represent the 56 signal diagrams of associated production of 
the top quark pair and Higgs boson. The remaining 53 signal diagrams can be 
obtained from those of Figs.~\ref{diags}(a), \ref{diags}(b) 
and \ref{diags}(c) by attaching the Higgs boson line of $hb\bar b$-vertex  
to another top or bottom quark line and interchanging the identical $b$- and 
$\bar b$-quarks in each of the 3 diagrams, and by interchanging the two gluon 
lines in Fig.~\ref{diags}(c).
The diagrams shown in the second row of Fig.~\ref{diags} are just a few
examples of the off resonance background contributions to associated production
of the top quark pair and Higgs boson. It should be noted that,
in the narrow width approximation, where the cross section of process
(\ref{gg8f}) factorizes into the cross section of process (\ref{pptth}) 
times the branching fractions of $t\to b u \bar{d}$, 
$\bar t\to b \mu^- \bar \nu_{\mu}$ and $h\to b\bar b$, there are only 
4 Feynman diagrams of process (\ref{gg8f}).

It has been shown in \cite{ggtth} that the LO differential distributions 
in rapidity and angles of the secondary lepton 
in the associated production of the top quark pair and Higgs boson 
in proton--proton collisions at the LHC are quite sensitive to 
modifications of the SM top--Higgs Yukawa coupling. 
In the present lecture, we will discuss the question to which extent 
the effects of anomalous couplings in the LO differential cross
sections and distributions depend on the choice of 
factorization scale in perturbative quantum chromodynamics.

\section{Non standard top--Higgs interaction}

Departures of the top--Higgs coupling from its SM form that
include corrections from dimension-six operators can be best 
parameterized in terms of the effective Lagrangian which, after eliminating
the redundant operators with the use equation of motion, 
has the following form \cite{aguilar}
\bea
\label{httcoupl}
\mathcal{L}_{t\bar t h}=-g_{t\bar th}\bar{t}\left(f+if'\gamma_5\right)t h,
\eea
where real couplings $f$ and $f'$ describe, respectively, scalar and 
pseudoscalar departures from the purely scalar top--Higgs 
interaction of SM that corresponds to $f=1$ and $f'=0$. 
$f$ and $f'$ are amongst least constraint couplings of the SM.
Currently only the following indirect constraints on $f$ at 95\% C.L. exist:
\bea
f\in [-1.2,-0.6]\cup[0.6,1.3]&&\qquad 
{\rm ATLAS}\;\cite{fATLAS}\nn\\
\label{indconstr}
f\in [0.3,1.0]\qquad\qquad\qquad\; &&\qquad 
{\rm CMS}\;\cite{fCMS}.\nn
\eea
They are derived from the process
of Higgs boson production through the gluon fusion, which is
dominated by the top-quark loop, and from the Higgs boson decay into 2 photons
that also receives a significant contribution from 
the top-quark loop.
However, the derivation relies on the assumptions that 
there is no new physical degrees of freedom
in the loops and that there are two
universal scale factors: one for all the Higgs boson Yukawa couplings to 
the SM fermion species and the other one for the Higgs boson couplings to 
electroweak gauge bosons.
The interval in the range of negative numbers is highly disfavoured,
as the opposite sign of the Higgs boson 
coupling to fermions with respect to its coupling to the gauge bosons 
is required in the Lagrangian for the unitarity and renormalizability 
of the theory \cite{unitarity} and vacuum stability \cite{stability}.

\section{Results}

Lagrangian (\ref{httcoupl}) has been implemented in {\tt carlomat} 
\cite{carlomat}, a general purpose program for Monte Carlo (MC) computation 
of lowest order cross sections. A new version of the program 
has already been made publicly available \cite{carlomat2}.
The cross section of 
\bea
\label{pp8f}
pp \;\ra\; b u \bar{d}\; \bar b \mu^- \bar \nu_{\mu} b \bar b
\eea
is computed by folding the cross section of the dominant hard scattering
gluon fusion process (\ref{gg8f}) with MSTW parton density functions 
(PDFs) \cite{MSTW} at the LO.

The complex mass scheme \cite{Racoon} is used in the computation 
and the initial physical input 
parameters are the same as in \cite{ggtth} except
for $\alpha_s(m_Z)=0.13939$ and the $b$-quark mass $m_b=4.75$~GeV, both
being transferred to {\tt carlomat} from the MSTW LO PDFs, and the Higgs boson 
width $\Gamma_h=7.1161$~MeV. 
The MC events of the associated production of the top quark pair and Higgs 
boson in process (\ref{pp8f}) are selected by identifying jets with their 
original partons and imposing cuts given by Eqs. (3.2)--(3.7) of \cite{ggtth},
with the $b\bar b$ invariant mass $m_{bb}^{\rm cut}=20$~GeV in Eq.~(3.7).

In order to test scale dependence of the LO differential cross sections and 
distributions of process (\ref{pp8f}), the factorization scale in MSTW PDFs 
is set to  $Q=q(2m_t+m_h)$, where the scale factor $q$ is chosen to be either 
$q=0.5$ or $q=2$. 

\begin{figure}[htb]
\vspace{100pt}
\begin{center}
\setlength{\unitlength}{1mm}
\begin{picture}(35,35)(0,0)
\includegraphics{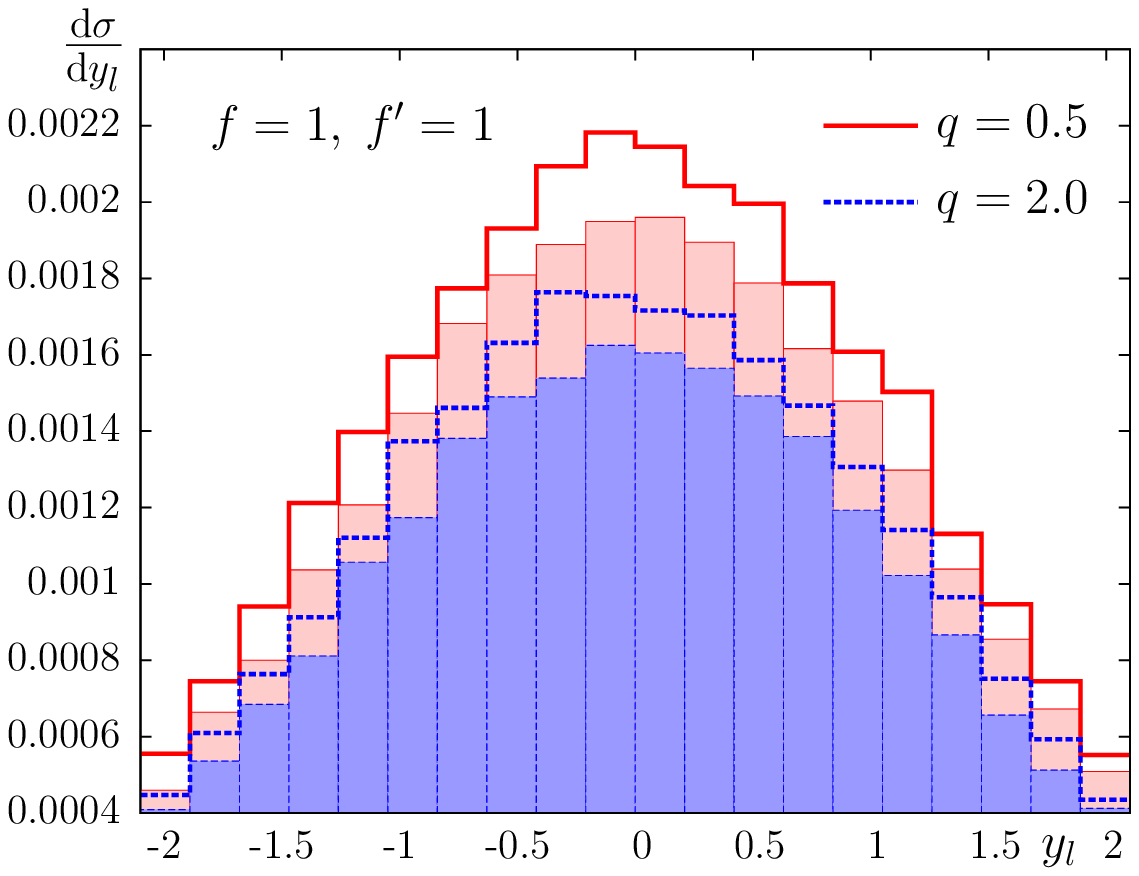}
\end{picture}
\hfill
\begin{picture}(35,35)(0,0)
\includegraphics{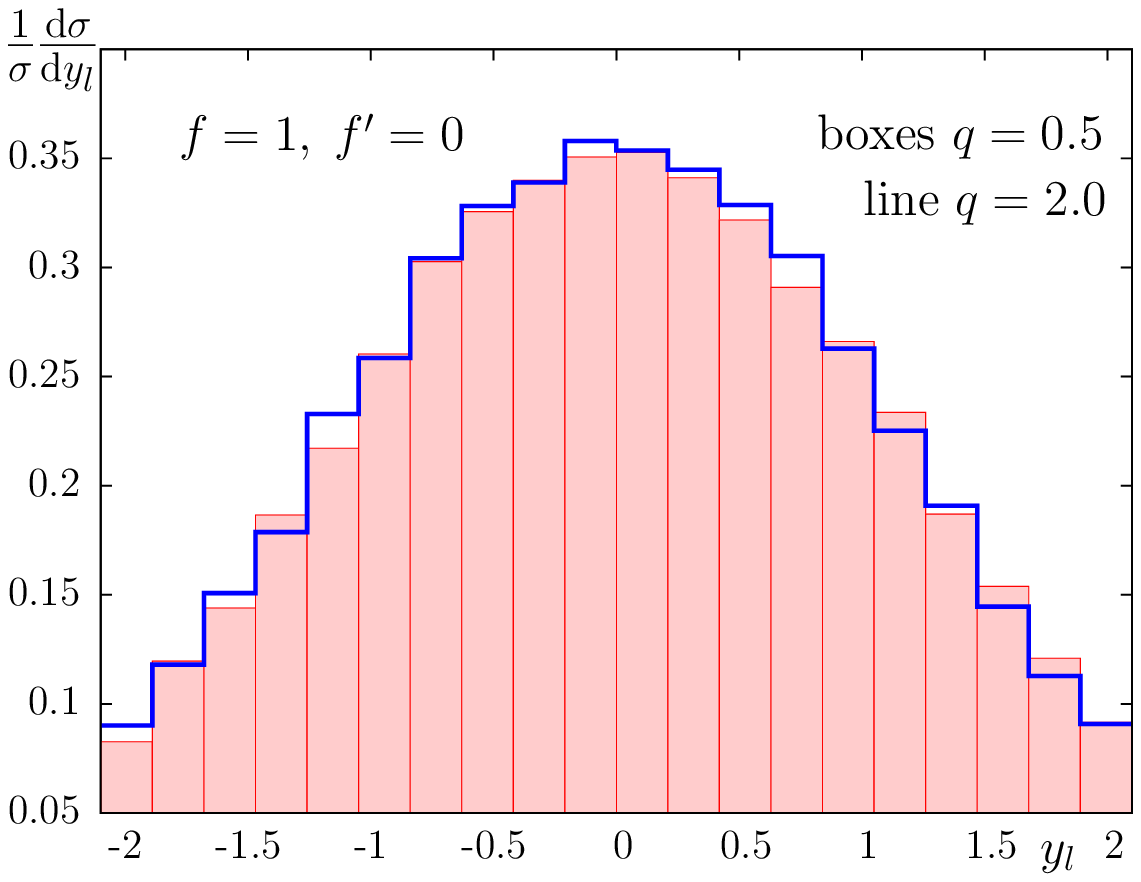}
\end{picture}\\[1.cm]
\begin{picture}(35,35)(0,0)
\includegraphics{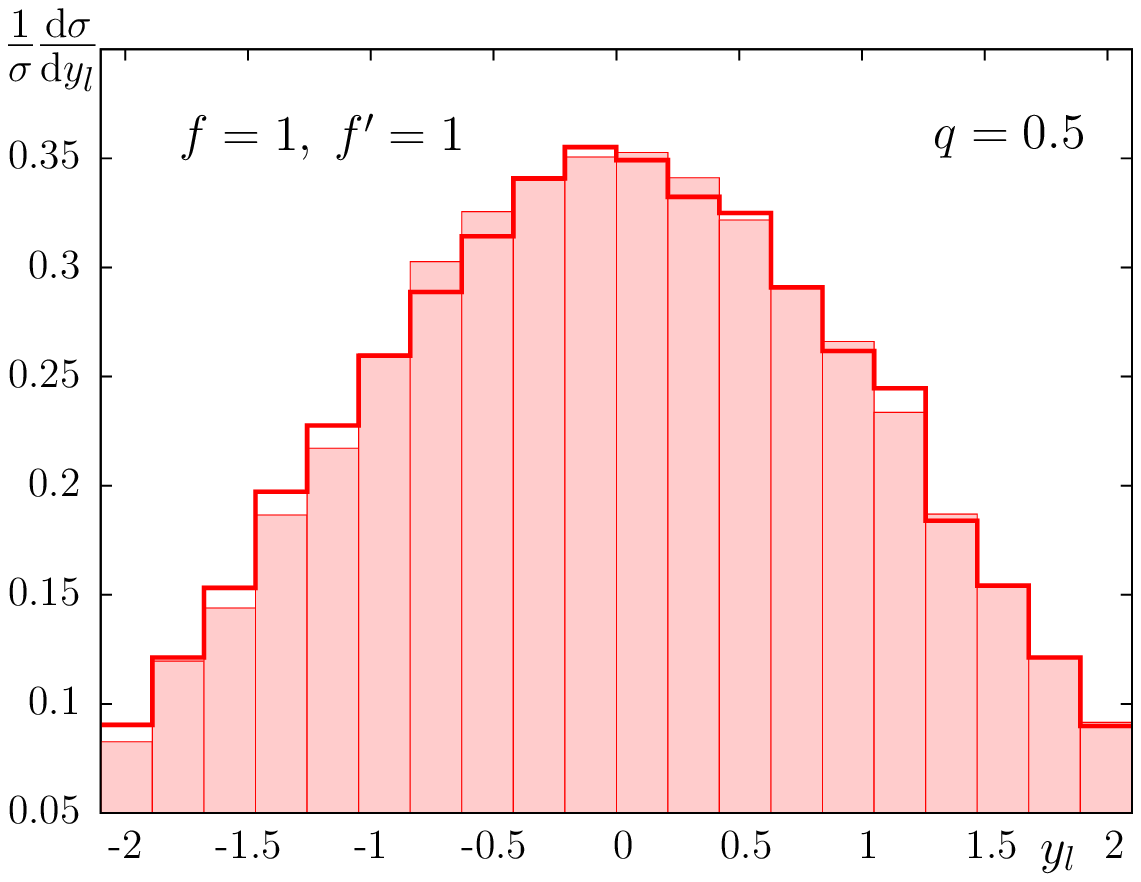}
\end{picture}
\hfill
\begin{picture}(35,35)(0,0)
\includegraphics{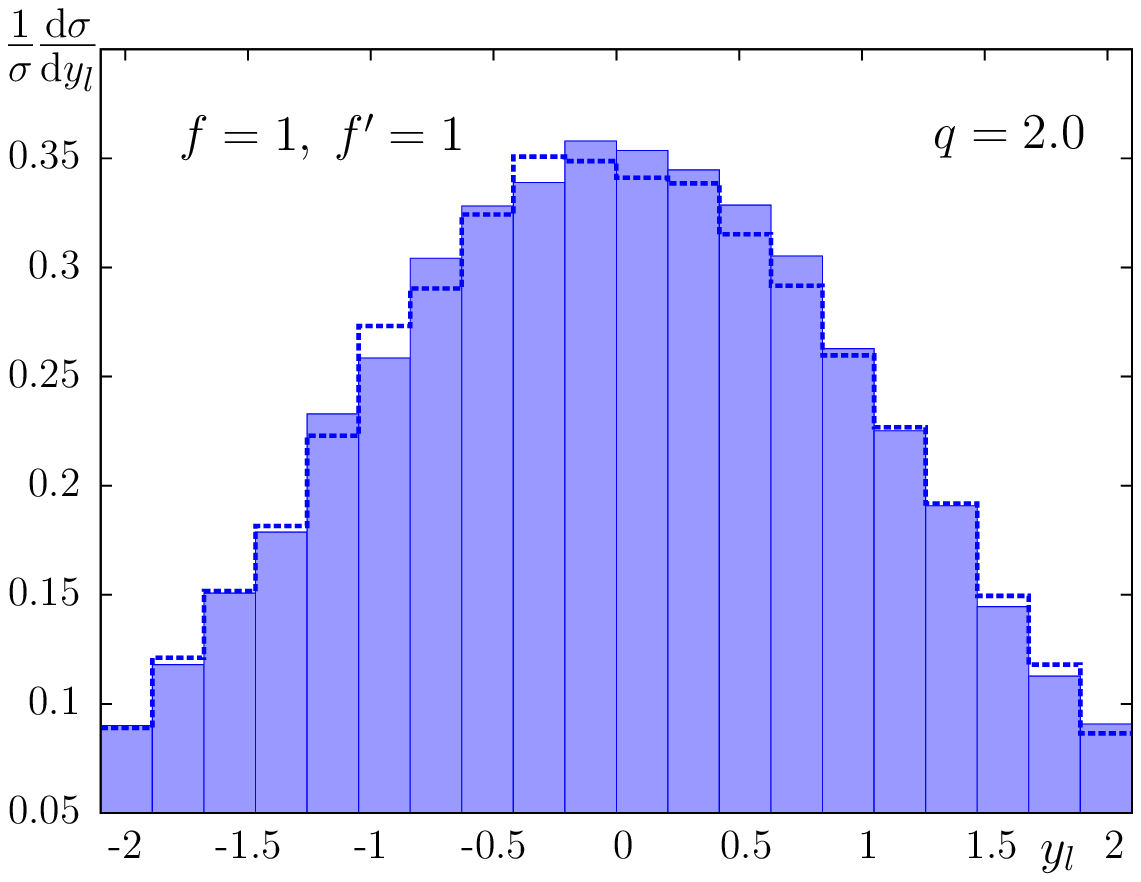}
\end{picture}
\end{center}
\vspace*{-2.cm}
\caption{Distributions in rapidity of the final state $\mu^-$ of process 
(\ref{pp8f}) in $pp$ collisions at $\sqrt{s}=14$~TeV.
}
\label{figrapl}
\end{figure}

The differential cross sections and normalized distributions
as functions of the rapidity of the final state $\mu^-$ of process
(\ref{pp8f}) in $pp$ collisions at $\sqrt{s}=14$~TeV are shown in 
Fig.~\ref{figrapl}. 
The plots in upper left hand side panel of Fig.~\ref{figrapl} show
the SM results, corresponding to $f=1$ and $f'=0$, for $q=0.5$ (boxes shaded 
in red) and $q=2$ (boxes shaded in blue) and the results for
$f=1$ and $f'=1$ that are plotted with lines: solid for $q=0.5$ and dashed 
for $q=2$. 
The scale dependence of the LO cross sections is substantial, 
as expected. It is to large extent reduced if the differential cross sections
are normalized. This can be seen in the upper right hand side panel,
where the SM results for rapidity distributions of $\mu^-$  
are plotted for $q=0.5$ (boxes) and $q=2$ (solid line).
The distributions computed with the anomalous choice
of couplings $f=1$ and $f'=1$, plotted with lines, are compared against 
the SM result, plotted with boxes, in the two lower panels of 
Fig.~\ref{figrapl} for $q=0.5$ (left panel) and $q=2$ (right panel). 
The effects of the anomalous couplings in the distributions are 
not big, as they are to large extent obscured by the off resonance
background contributions to the associated production of the top quark pair 
and Higgs boson, which was shown in \cite{ggtth}.

\begin{figure}[htb]
\vspace{100pt}
\begin{center}
\setlength{\unitlength}{1mm}
\begin{picture}(35,35)(0,0)
\includegraphics{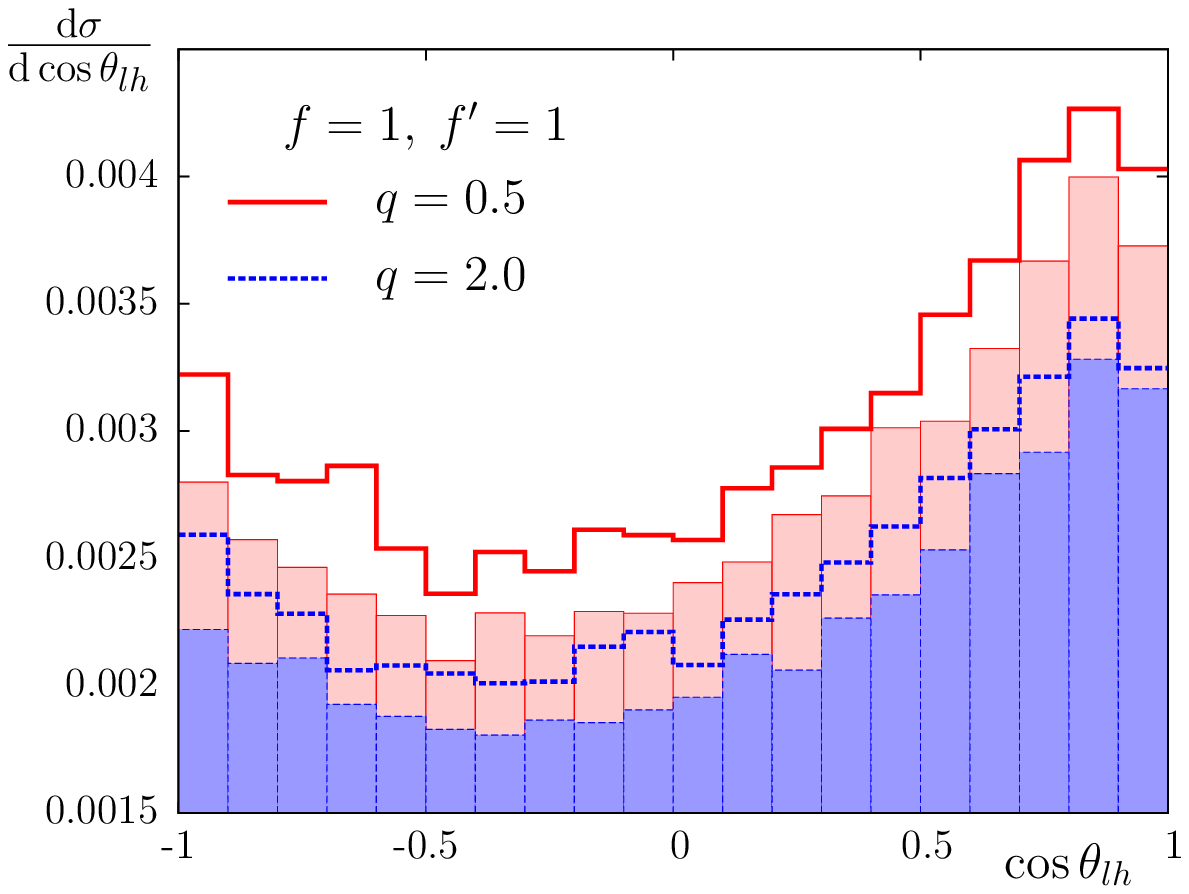}
\end{picture}
\hfill
\begin{picture}(35,35)(0,0)
\includegraphics{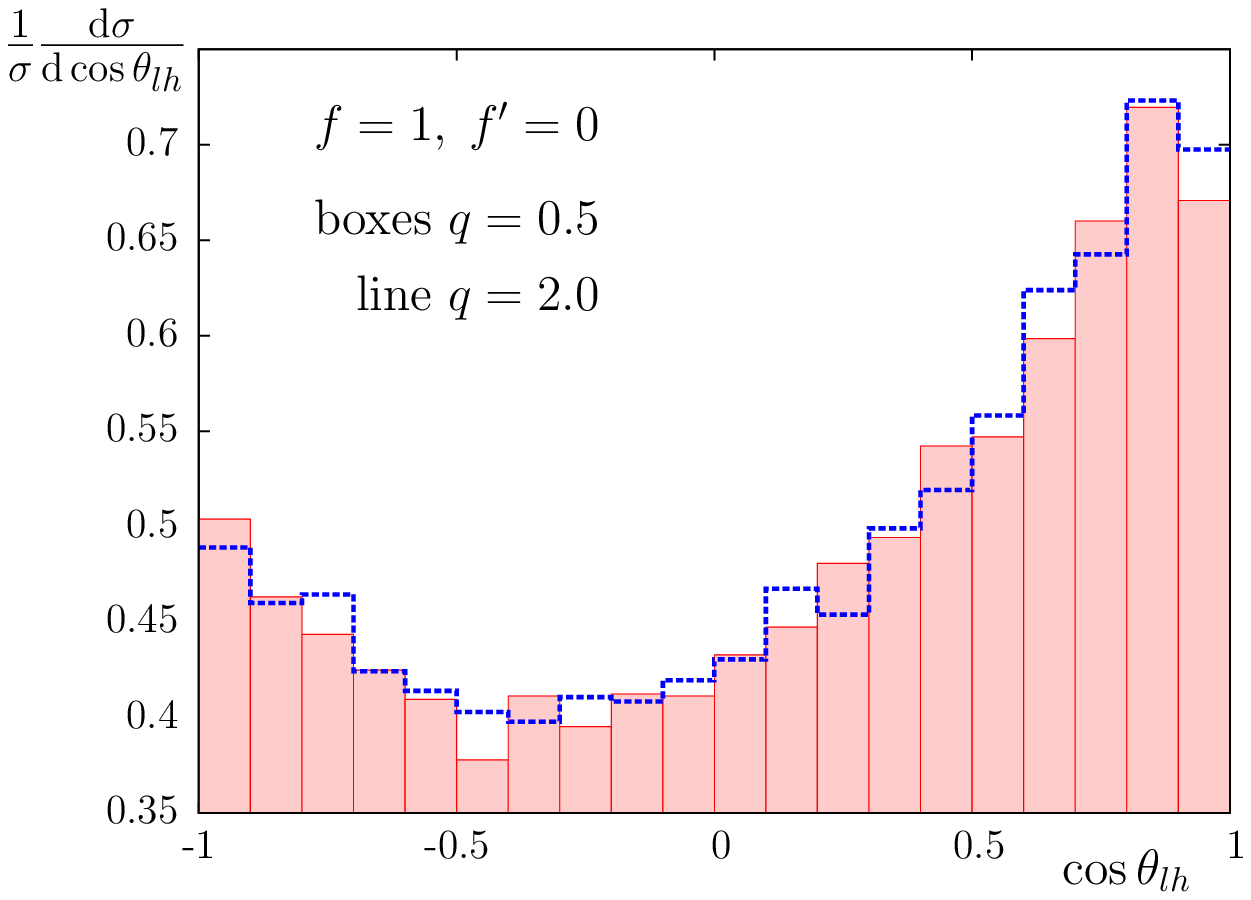}
\end{picture}\\[1.cm]
\begin{picture}(35,35)(0,0)
\includegraphics{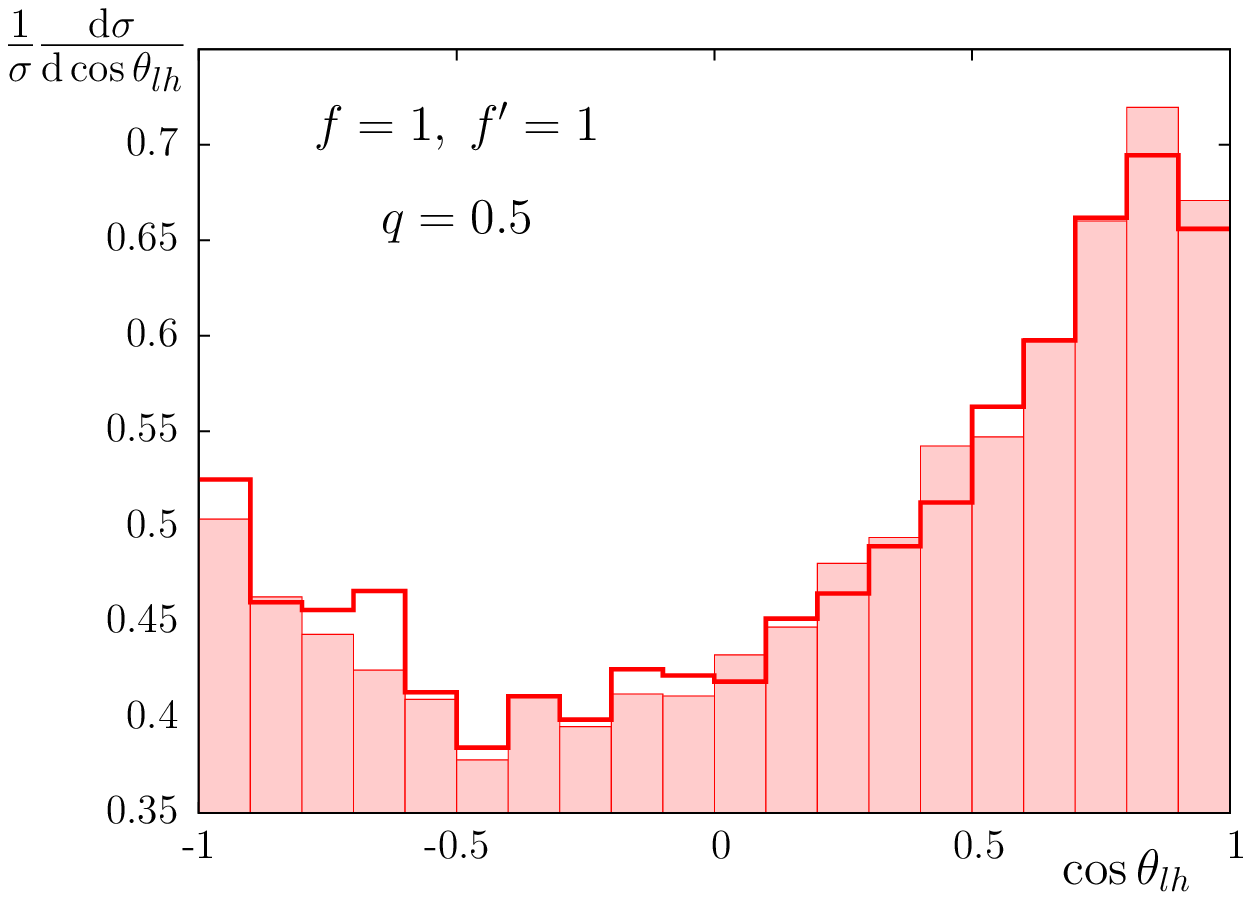}
\end{picture}
\hfill
\begin{picture}(35,35)(0,0)
\includegraphics{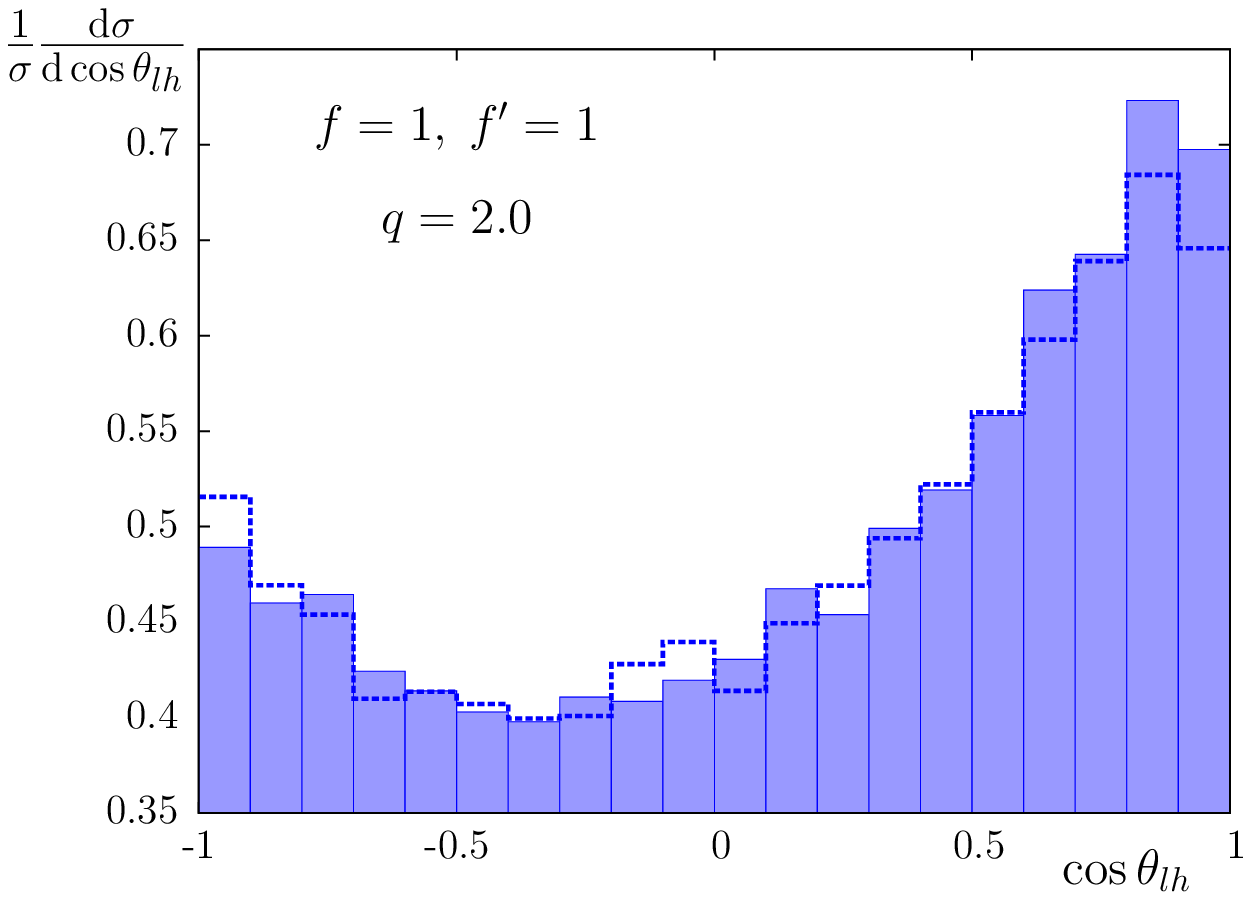}
\end{picture}
\end{center}
\vspace*{-2.cm}
\caption{Distributions in cosine of the angle between $\mu^-$ and the
reconstructed Higgs boson momentum of process 
(\ref{pp8f}) in $pp$ collisions at $\sqrt{s}=14$~TeV.
}
\label{figcth}
\end{figure}

Practically the same observations hold for the differential cross sections 
and normalized distributions as functions of the angle between
$\mu^-$ and the reconstructed momentum of the Higgs boson of process
(\ref{pp8f}) in $pp$ collisions at $\sqrt{s}=14$~TeV that are shown in 
Fig.~\ref{figcth}. 

\begin{figure}[htb]
\vspace{100pt}
\begin{center}
\setlength{\unitlength}{1mm}
\begin{picture}(35,35)(0,0)
\includegraphics{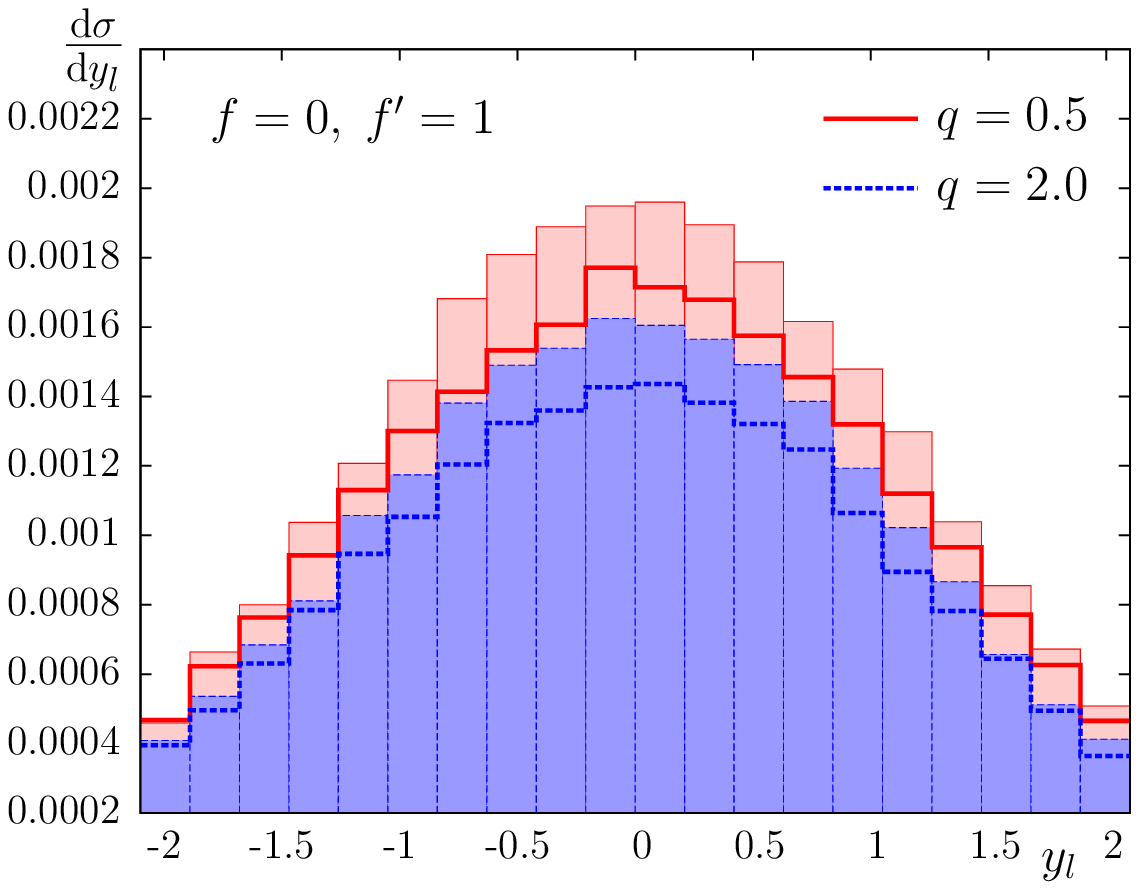}
\end{picture}
\hfill
\begin{picture}(35,35)(0,0)
\includegraphics{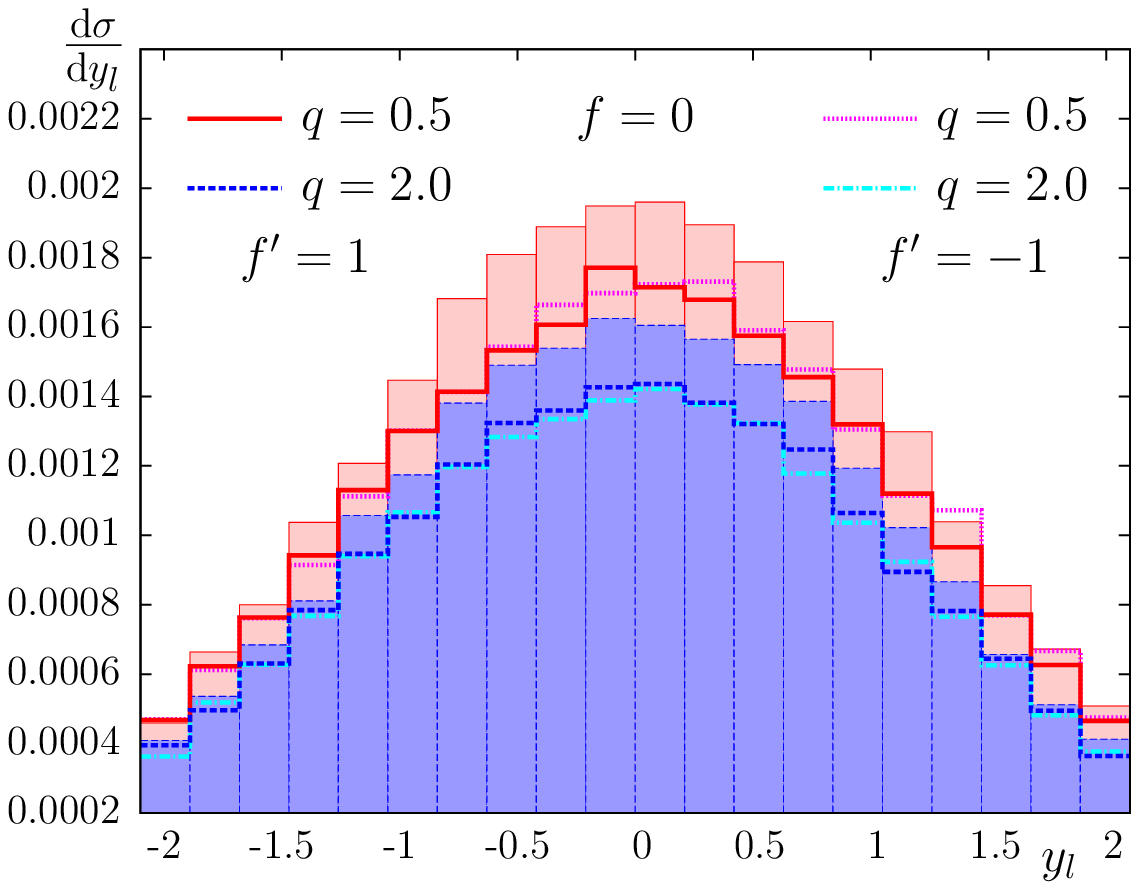}
\end{picture}
\end{center}
\vspace*{-2.cm}
\caption{Distributions in rapidity of the final state $\mu^-$ of process
(\ref{pp8f}) in $pp$ collisions at $\sqrt{s}=14$~TeV.
}
\label{figrapl1}
\end{figure}

It would be interesting to see whether or not the differential cross sections 
of process (\ref{pp8f}) are sensitive to a sign of $f'$. For the sake 
clarity, let us assume $f=0$ which, despite being beyond limits of 
(\ref{indconstr}), is still not excluded by direct constraints.
In Fig.~\ref{figrapl1}, the differential cross sections as functions 
of the rapidity of the final 
state $\mu^-$ of process (\ref{pp8f}) at $\sqrt{s}=14$~TeV 
for two other anomalous combinations of couplings: $f=0$ and $f'=\pm 1$
with $q=0.5$ and $q=2$ are plotted with lines together with 
the corresponding SM results that are plotted with boxes shaded 
in red for $q=0.5$ and in blue for $q=2$.
The left panel shows the results for $f=0$ and $f'=1$
while the right panel shows the results for both $f=0$ and $f'=1$,
and $f=0$ and $f'=-1$.
It can be seen that the cross sections show rather little  sensitivity 
to a sign of the anomalous pseudoscalar coupling.

\section{Summary and outlook}

The factorization scale dependence of the 
LO differential cross sections and distributions in the process of 
associated production of the top quark 
pair and Higgs boson at the LHC in the presence of anomalous top--Higgs
coupling has been discussed.
The substantial scale dependence of the LO cross sections is to large extent 
reduced if the corresponding normalized distributions are considered.
It has also been shown that the differential cross section as a function
of the rapidity of the final 
state $\mu^-$ of process (\ref{pp8f}) at $\sqrt{s}=14$~TeV 
is practically not sensitive
to a sign of the anomalous pseudoscalar coupling.

Process (\ref{pp8f}) may be affected by many other possible 
deviations from the SM couplings that have not been discussed in this lecture, 
where we have focused just on the effects of the anomalous $t\bar t h$ 
interaction
on the distributions of the secondary lepton. However, some of the deviations
could be easily included in the discussion as they have been already 
implemented in {\tt carlomat} \cite{carlomat2}. This holds in particular 
for the anomalous $Wtb$ coupling whose effects on the process of top quark 
pair production in hadronic collisions was studied in \cite{wtbafb} and
\cite{wtblhc}.

Acknowledgements: This project was supported in part with financial resources 
of the Polish National Science Centre (NCN) under grant decision 
number DEC-2011/03/B/ST6/01615 and by the Research Executive 
Agency (REA) of the European Union under the Grant Agreement number 
PITN-GA-2010-264564 (LHCPhenoNet).


\begin{thebibliography}{99}
\bibitem{tthCMS} CMS Collaboration, JHEP {\bf 05} (2013) 145
[arXiv:1303.0763].
\bibitem{ggtth} K. Ko\l odziej, J. High Energy Phys. {\bf 1307}
      (2013) 083, [arXiv:1303.4962 [hep-ph]].
\bibitem{aguilar} J.A. Aguilar-Saavedra, Nucl. Phys. {\bf B821} (2009) 215
[arXiv:0904.2387].
\bibitem{fATLAS} ATLAS Collaboration, {\em Combined coupling measurements 
of the Higgs-like boson with the
ATLAS detector using up to $25\;{\rm fb}^{-1}$ of proton-proton collision data},
ATLAS-CONF-2013-034, March 13, 2013.
\bibitem{fCMS} CMS Collaboration, {\em Observation of a resonance with a mass 
near 125 GeV in the search for the Higgs boson in $pp$ collisions at 
$\sqrt{s} = 7$~TeV and 8~TeV}, CMS-PAS-HIG-12-020, July 2012.
\bibitem{unitarity} T. Appelquist, M. S. Chanowitz, 
Phys. Rev. Lett. {\bf 59} (1987) 2405 [Erratum-ibid. {\bf 60} (1988) 1589].
\bibitem{stability} M. Reece,New J. Phys. {\bf 15} (2013) 043003 
               [arXiv:1208.1765[hep-ph]].
\bibitem{carlomat} K. Ko\l odziej, Comput. Phys. Commun. {\bf 180} (2009) 
                   1671;\\ 
                   K. Ko\l odziej, Acta Phys. Polon. {\bf B42} (2011) 2477.
\bibitem{carlomat2} K. Ko\l odziej, Comput. Phys. Commun. (2013) 
                    doi: 10.1016/j.cpc.2013.08.023 [arXiv:1305.5096[hep-ph]];
                    program available from
                    CPC Program Library, or from 
                    {\tt http://kk.us.edu.pl/carlomat.html}.
\bibitem{MSTW} A.D. Martin, W.J. Stirling, R. S. Thorne, G. Watt,
  Eur. Phys. J. {\bf C63} (2009) 189.
\bibitem{Racoon} A. Denner, S. Dittmaier, M. Roth, D. Wackeroth,
                 Nucl. Phys. {\bf B560} (1999) 33 and
                 Comput. Phys. Commun. {\bf 153} (2003) 462.
\bibitem{wtbafb} K. Ko\l odziej, Phys. Lett. {\bf B710} (2012), 671
                 [arXiv:1110.2103[hep-ph]].
\bibitem{wtblhc} K. Ko\l odziej, Acta Phys. Pol. {\bf B44} (2013) 1775
               [arXiv:1212.6733[hep-ph]].
\end{thebibliography}
\end{document}